\documentclass[twocolumn,a4paper,accepted=2021-04-12]{quantumarticle}

\pdfoutput=1
\usepackage[utf8]{inputenc}
\usepackage[usenames,dvipsnames]{xcolor}
\usepackage[english]{babel}
\usepackage[T1]{fontenc}
\usepackage{amsmath,amssymb,amsfonts,microtype,ragged2e}

\usepackage{appendix,hyphenat,mathtools,dsfont,bm,bbm,amsthm,tikz,pgfplots}
\usepackage[makeroom]{cancel}
\usepackage[numbers,sort&compress]{natbib}

\definecolor{mygrey}{gray}{0.35}
\definecolor{myblue}{rgb}{0.2,0.2,0.8}
\definecolor{myzard}{cmyk}{0,0,0.05,0}
\definecolor{mywhite}{rgb}{1,1,1}
\definecolor{myred}{rgb}{0.9,0.1,0.}
\definecolor{dgreen}{rgb}{0.0, 0.5, 0.0}
\usepackage[colorlinks=true,citecolor=myblue,linkcolor=myblue,urlcolor=myblue]{hyperref}
\usepackage[capitalize]{cleveref}
\usepackage{bookmark}

\newcommand{\ket}[1]{\lvert#1\rangle} 
\newcommand{\bra}[1]{\langle#1\rvert} 

\makeatletter
\def\mbig{\bBigg@{1.2}}
\def\mbigl#1{\mathopen{\bBigg@{1.2}#1}}
\def\mbigr#1{\mathclose{\bBigg@{1.2}#1}}
\renewcommand*\env@matrix[1][*\c@MaxMatrixCols c]{%
   \hskip -\arraycolsep%
   \let\@ifnextchar\new@ifnextchar%
   \array{#1}%
}
\makeatother

\newcommand*\bgcol[2]{{%
   \ifmmode%
      \mathchoice%
         {\colorbox{#1}{$\displaystyle#2$}}%
         {\colorbox{#1}{$\textstyle#2$}}%
         {\colorbox{#1}{$\scriptstyle#2$}}%
         {\colorbox{#1}{$\scriptscriptstyle#2$}}%
   \else%
      \colorbox{#1}{#2}%
   \fi%
}}

\def\d{\mathop{}\!\mathrm d\mathchoice{}{}{\kern-.09em}{\kern-.09em}}



\pgfplotsset{compat=1.14}

\begin{document}
\title{On the connection between microscopic description and memory effects in open quantum system dynamics}

\author{Andrea Smirne}
\affiliation{Dipartimento di Fisica “Aldo Pontremoli”, Università degli Studi di Milano, via Celoria 16, 20133 Milan, Italy}
\affiliation{Istituto Nazionale di Fisica Nucleare, Sezione di Milano, via Celoria 16, 20133 Milan, Italy}
\author{Nina Megier}
\affiliation{Dipartimento di Fisica “Aldo Pontremoli”, Università degli Studi di Milano, via Celoria 16, 20133 Milan, Italy}
\affiliation{Istituto Nazionale di Fisica Nucleare, Sezione di Milano, via Celoria 16, 20133 Milan, Italy}
\author{Bassano Vacchini}
\affiliation{Dipartimento di Fisica “Aldo Pontremoli”, Università degli Studi di Milano, via Celoria 16, 20133 Milan, Italy}
\affiliation{Istituto Nazionale di Fisica Nucleare, Sezione di Milano, via Celoria 16, 20133 Milan, Italy}

\begin{abstract}
The exchange of information between an open quantum system and its environment
allows us to discriminate among different kinds of dynamics,
in particular detecting 
memory effects to characterize non-Markovianity. Here, we investigate the role played by
the system-environment correlations and the environmental evolution in the 
flow of information.  First, we derive general conditions ensuring
that two generalized dephasing microscopic models of the global
system-environment evolution result exactly in the same open-system
dynamics, for any initial state of the system. Then, we use the trace
distance to quantify the distinct contributions to the information
inside and outside the open system in the two models.  Our analysis
clarifies how the interplay between system-environment correlations
and environmental-state distinguishability can lead to the same
information flow from and toward the open system, despite significant
qualitative and quantitative differences at the level of the global
evolution.
\end{abstract}

\maketitle

  \section{Introduction}
\label{sec:intro}

Whenever we want to describe the time evolution of a quantum system taking
into account the effects of the surrounding environment, we can rely on the tools 
provided by the theory of open quantum systems \cite{Breuer2002,Rivas2012}. The latter, in fact, allows us to model physical phenomena, such as dissipation and decoherence, that are inherently associated with the open-system nature of the quantum system at hand. Generally speaking, quantities that would be 
conserved under a closed-system unitary evolution will rather change in time as a consequence of the action of the environment. In somehow more abstract terms, 
the interaction between a quantum system and its environment induces a mutual exchange of information, which would be prevented if the system were closed.

Besides discriminating closed-system and open-system evolutions, 
the exchange of information between an open quantum system and its environment also provides us with a powerful way to distinguish different open quantum system dynamics,
associated with qualitatively and quantitatively distinct behaviours.
In certain dynamics, the information flows unidirectionally from the open system to the environment, so that, once leaked out of the open system, it is irremediably lost.  
In other dynamics, instead, there is a bi-directional flow of information, implying that some information previously flown from the reduced system to the environment can later follow the opposite path; in other terms, 
the environment, as well as the system-environment correlations, can act as a memory storage, giving back to the open system some information that was 
previously contained in it.  
Relying on this intuition, the backflow of information to a reduced
system can be regarded as the distinctive sign 
of the presence of memory in its evolution. This, in turn, leads to the identification of
open-system dynamics having a two-fold exchange of information between the open
system and the environment with quantum non-Markovian dynamics, that is, dynamics where memory effects cannot be neglected (analogously to the corresponding notion
for classical stochastic processes \cite{Feller1971,Vacchini2011a,Vacchini2012a,Rivas2014,Breuer2016}).  
This is precisely the route that has been established in \cite{Breuer2009,Laine2010},
where the picture above has been defined in rigorous terms by means of the trace distance,
used as a quantifier of quantum-state distinguishability \cite{Fuchs1999}:
The variations in time of the trace distance detect the direction of the information flow between the open
system and the environment and then the Markovian or non-Markovian nature 
of the corresponding dynamics.

Despite the relevant theoretical \cite{Rivas2014,Breuer2016,Li2017,Vega2017,Li2019} and experimental \cite{Liu2011,Bernardes2016,Cialdi2017,Haase2018,Wittemer2018,Li2020} progresses in understanding 
the differences between Markovian and non-Markovian quantum dynamics, several
key questions remain to be addressed. In particular, it would be desirable
to connect the possible occurrence of memory
effects in open system dynamics with general features of the underlying microscopic description
of the open system, its environment and their interaction. 
Within the trace distance approach, it is possible to ascribe
any backflow of information towards the open system 
to either the generation of system-environment correlations,
or changes in the environmental state due to the interaction with the open system, or
both \cite{Laine2010b,Breuer2016}. Even more, quantitative links
between the trace distance variations and the influence of both system-environmental correlations
and environmental changes, as measured via the trace distance, have
been derived \cite{Mazzola2012,Smirne2013,Campbell2019}, and a similar
result has recently been obtained also for entropic distinguishability
quantifiers \cite{Megier2021a}.
In addition to their quantitative content, these links suggest further evidence that the possible quantum
nature of the system-environment correlations, in terms of the presence of entanglement \cite{Bengtsson2006} or non-zero discord \cite{Ollivier2001,Henderson2001,Modi2012}, does not  
play any special role in producing memory effects, compared to mere
classical correlations \cite{Pernice2011,Pernice2012}.
Indeed, the key point is that the state of an open quantum system, and then any information-content
associated with it, is the result of an average over the environmental degrees of freedom, 
mathematically described by the partial trace \cite{Breuer2002,Rivas2012}.
As a consequence, different system-environment correlations and
environmental states might well result in exactly the same reduced system
evolution. 

Role and relevance of correlations with ancillary degrees of
freedom in the characterization of non-Markovian dynamics has also
been the object of many recent investigations
\cite{DeSantis2019a,Kolodynski2020,DeSantis2020a,DeSantis2020b}. 
In the present contribution, we instead concentrate on the role of the
correlations between system and environment arising due to the
microscopic interaction. The latter, together with the impact of the
interaction on the environment, should be the ultimate cause of memory effects.
More specifically, we investigate by means of the trace distance
to what extent different evolutions of the information lying outside the open system -- being
in the system-environment correlations or in the environmental state --
can lead to the same information flow from and toward the open quantum system.
This analysis will help clarify the non-trivial interplay between the features of the global
evolution that can provoke
non-Markovian open system dynamics. 

We first consider the generalized pure-dephasing dynamics \cite{Breuer2002} 
of a $d$-dimensional open quantum system interacting with a generic environment
and, relying on the exact analytical solution, we derive general conditions ensuring the equivalence between two open system dynamics. These dynamics result from two distinct microscopical models, for which the type of the environment, the initial environmental state and the interaction between the system and the environment may differ. After moving
to the simplest scenario involving a two-level system and two-level environments,
we show that the reduced system dynamics can coincide even though in one case the global
state is classically correlated, while in the other it is (almost) always entangled (see Fig.\ref{fig:sk}), 
and even maximally entangled at isolated instants of time.
By means of the bound derived in \cite{Laine2010b}, we compare 
the strength of the system-environment correlations and environmental changes 
in the two global evolutions, 
showing that relevant qualitative and quantitative differences among them can still result in the same
exchange of information between the open system and the environment, and thus in the same
non-Markovian behavior.

The rest of the paper is organized as follows. In Sec.\ref{sec:info-exchange}, we recall the features
of the trace distance characterization of quantum non-Markovianity that are relevant for our analysis. 
In Sec.\ref{sec:different-microscopic}, we derive explicit conditions on the environmental initial states and interaction operators guaranteeing that different generalized pure dephasing microscopic models lead to the same open system dynamics.
Sec.\ref{sec:secorr} contains the main part of our paper, where the qualitative and quantitative differences
between the system-environment correlations and environmental states in the two models are studied 
in relation with their influence on the occurrence of memory effects, as signaled by an increase
of the trace distance. Finally, the conclusions and possible outlooks are given in Sec.\ref{sec:concl}.

\section{System-environment information exchange and quantum Markovianity}
\label{sec:info-exchange}
In order to fix the notation and introduce the notions that will be relevant 
for the rest of the paper, we start by briefly recalling the mathematical characterization
of quantum Markovianity in terms of the trace distance, along with its 
physical interpretation in connection with the exchange of information 
between an open quantum system and its environment \cite{Breuer2009,Laine2010,Breuer2016}.

Given the Hilbert space $\mathcal{H}_S$ associated with an open quantum system
and the set of statistical operators $\mathcal{S}(\mathcal{H}_S)$, i.e., the self-adjoint, positive, trace-one
operators acting on $\mathcal{H}_S$, we denote as $\rho_S(t)\in \mathcal{S}(\mathcal{H}_S)$ 
the state of the open system at time $t$. Under the assumptions
that the open system and the environment can be
treated overall as a closed system and that they are uncorrelated at the
initial time $t_0=0$, i.e., the initial global state
is $\rho_{SE}(0) = \rho_S(0)\otimes \rho_E(0)$ with $\rho_E(0)$
a fixed environmental state (within the set $\mathcal{S}(\mathcal{H}_E)$
of statistical operators on $\mathcal{H}_E$), 
the state $\rho_S(t)$ is given by the completely
positive trace preserving (CPTP) map $\Lambda(t)$ defined as \cite{Breuer2002,Rivas2012}
\begin{eqnarray}\label{eq:lambda}
\rho_S(t) &=& \Lambda(t)[\rho_S(0)] \nonumber\\
&:=&\mbox{tr}_E\left\{U_{SE}(t)(\rho_S(0)\otimes \rho_E(0))U^{\dag}_{SE}(t)\right\}.
\end{eqnarray}
Here and in the following, $\mbox{tr}_E$ ($\mbox{tr}_S$) denotes the partial trace over the environmental
(open system) degrees of freedom
and $U_{SE}(t)$ is the unitary operator describing the global closed-system evolution
from the initial time to the time $t$.

The family of CPTP maps at the different times, $\left\{\Lambda(t)\right\}_{t\geqslant 0}$,
fixes the open system dynamics and
encloses all the predictions related
with measurements performed on the open system, at any single time $t$ and for any initial condition 
$\rho_S(0)$\footnote{In general, instead, the family of CPTP maps 
is not enough to fully characterize the 
statistics associated with multi-time measurements, for which different mathematical objects are needed; suitable
notions of quantum Markovianity can be defined also for such objects and are indeed not
equivalent to the notions referred to the open-system dynamics \cite{Breuer2016,Li2017,Pollock2018,Milz2019,Smirne2019,Milz2020,Milz2020b}.}.
As a consequence, the different features of open system dynamics are preferably
formulated in terms of properties of the corresponding families of maps. Indeed, this is the case
also for quantum Markovianity, which aims at introducing
the notion of memoryless evolutions in the quantum realm, analogously to what 
happens for classical stochastic processes \cite{Feller1971,Breuer2002}. 
Among the different, and not necessarily equivalent, definitions of Markovian quantum dynamics 
\cite{Wolf2008,Rivas2010,Lu2010,Rivas2014,Chruscinski2014,Hall2014,Buscemi2016,Li2017,Megier2017,DeSantis2019a,Kolodynski2020,Jahromi2020}, the one based on the trace distance \cite{Breuer2009,Laine2010,Breuer2016}
stems from a quantitative definition of memory effects, 
linked to the information exchange between the system of interest and its environment.

The trace distance between two quantum states $\rho^1$ and $\rho^2$, which is defined as
\begin{equation}
D(\rho^1,\rho^2) = \frac{1}{2} \left\|\rho^1-\rho^2\right\|_1 = \frac{1}{2}\sum_i |x_i|
\end{equation}
with $\| \cdot\|_1$ the trace norm and hence $x_i$ the eigenvalues of $\rho^1-\rho^2$,
quantifies their distinguishability \cite{Fuchs1999}, that is, the ability to discriminate
between $\rho^1$ and $\rho^2$ if it is known that one of the two states has been prepared with probability
$1/2$; note that a more general quantifier of distinguishability can be introduced, including a possibly biased
probability of preparation \cite{Chruscinski2011,Wissmann2015,Breuer2016}.
Now, if we consider the evolution of the trace distance $D(\rho_S^1(t), \rho^2_S(t))$ between
two open system states $\rho_S^1(t)$ and $\rho^2_S(t)$,
evolved from two different initial
conditions $\rho_S^1(0)$ and $\rho^2_S(0)$ via Eq.(\ref{eq:lambda}),
the decrease in time of $D(\rho_S^1(t), \rho^2_S(t))$ can be traced back to a loss of information
from the open system, due to the interaction
with the environment. On the same footing, an increase in the trace distance means
that some information is flowing back to the open system, leading to an increased capability
to discriminate between the two possible states by performing measurements on the reduced system
only. Such a backflow of information is precisely what is identified as memory effect in the definition
of quantum Markovianity introduced in \cite{Breuer2009,Laine2010}.
Following that definition, we say that non-Markovian
dynamics are those where there is at least a pair of initial states and a time interval $[s,t]$, with
$t\geq s$,
such that
\begin{equation}\label{eq:variation}
\Delta_S(t,s) := D(\rho_S^1(t), \rho^2_S(t)) - D(\rho_S^1(s), \rho^2_S(s))
\end{equation}
is larger than zero. Importantly, by integrating the time derivative of the trace distance 
over the time intervals where 
$\Delta_S(t,s)>0$ and optimizing the integral 
over the pairs of initial conditions, one can introduce
a measure of non-Markovianity that is univocally associated
with the family of CPTP dynamical maps. 
At the same time,
the detection of an increase in the trace distance for a
single pair of initial states and interval of time
is enough to witness the non-Markovianity of the dynamics, 
which then does not call for the full reconstruction of the dynamical map, nor for an explicit
microscopic model of the underlying system-environment interaction \cite{Breuer2009,Laine2010}.

Besides the rigorous definition of quantum (non)-Markovian dynamics
rooted in the information exchange between the open system and the environment, 
the trace distance
also provides us with a clear physical picture motivating the possible occurrence
of memory effects. 
The contractivity of the trace distance under CPTP maps, along with the triangular inequality, 
allows us to upper bound the trace distance variation 
in Eq.(\ref{eq:variation}) via \cite{Laine2010b,Breuer2016}
\begin{equation}
  \label{eq:bound}
  \Delta_S(t,s) \leqslant I_{SE}(s)
\end{equation}
with
\begin{eqnarray}\label{eq:bound2}
I_{SE}(s):= && D(\rho^1_{E}(s), \rho^2_E(s)) + D(\rho^1_{SE}(s), \rho^1_S(s) \otimes \rho^1_E(s))\nonumber\\
&&
+ D(\rho^2_{SE}(s), \rho^2_S(s) \otimes \rho^2_E(s)); 
\end{eqnarray}
here $\rho_S(s) = \mbox{tr}_E\left\{\rho_{SE}(s)\right\}$ ($\rho_E(s)=\mbox{tr}_S\left\{\rho_{SE}(s)\right\}$)
denotes the reduced system (environmental) state at time $s$
obtained from the global state $\rho_{SE}(s)\in \mathcal{S}(\mathcal{H}_{SE})$. 
The terms at the right hand side (r.h.s.) of the previous inequality describe, respectively,
the difference between the two environmental states
$\rho^1_{E}(s)$ and $\rho^2_{E}(s)$
at time $s$ and the total correlations in the two global states $\rho^1_{SE}(s)$
and $\rho^2_{SE}(s)$; the labels $1$ and $2$ refer to the two different initial reduced system states
$\rho^1_S(0)$ and $\rho^2_S(0)$.
Crucially, Eq.(\ref{eq:bound}) relates the trace distance between open system states
with quantities that refer to the system-environment correlations
and the environmental state, and that are thus associated 
with some information lying outside the open system itself.
On the one hand, this provides us with an explanation of the physical origin of 
memory effects in quantum dynamics, as an increase in the trace distance at time $t$,
$\Delta_S(t,s)>0$, is necessarily linked to the presence, at the previous time $s$, 
of system-environment correlations
and/or to differences in the environmental states due to the different
initial conditions.
On the other hand, in a complementary way, we can use the bound in Eq.(\ref{eq:bound})
as a starting point to gain some quantitative information about the system-environment correlations
and the changes in the environment established by the interaction, 
via measurements performed on the reduced
system only \cite{Mazzola2012,Smirne2013,Campbell2019}.

\section{Locally indistinguishable microscopic models}
\label{sec:different-microscopic}

The correlations between an open system and its environment and the dependence of the environmental
state on the reduced system initial condition necessarily feed
any backflow of information to the open system. 
However, one should keep in mind that the reduced system state 
is related to the global state via the partial trace
in Eq.(\ref{eq:lambda}), which unavoidably washes out the details about the global dynamics
that do not have an impact on the open-system evolution.
To fully appreciate in what respect system-environment correlations and environmental states
affect the flow of information from and toward the reduced system, it is thus important
to understand to what extent different global evolutions can lead to similar, or even to the same open-system
dynamics.
\begin{figure}[ht!]
\begin{center}
	\includegraphics[width=0.45\textwidth]{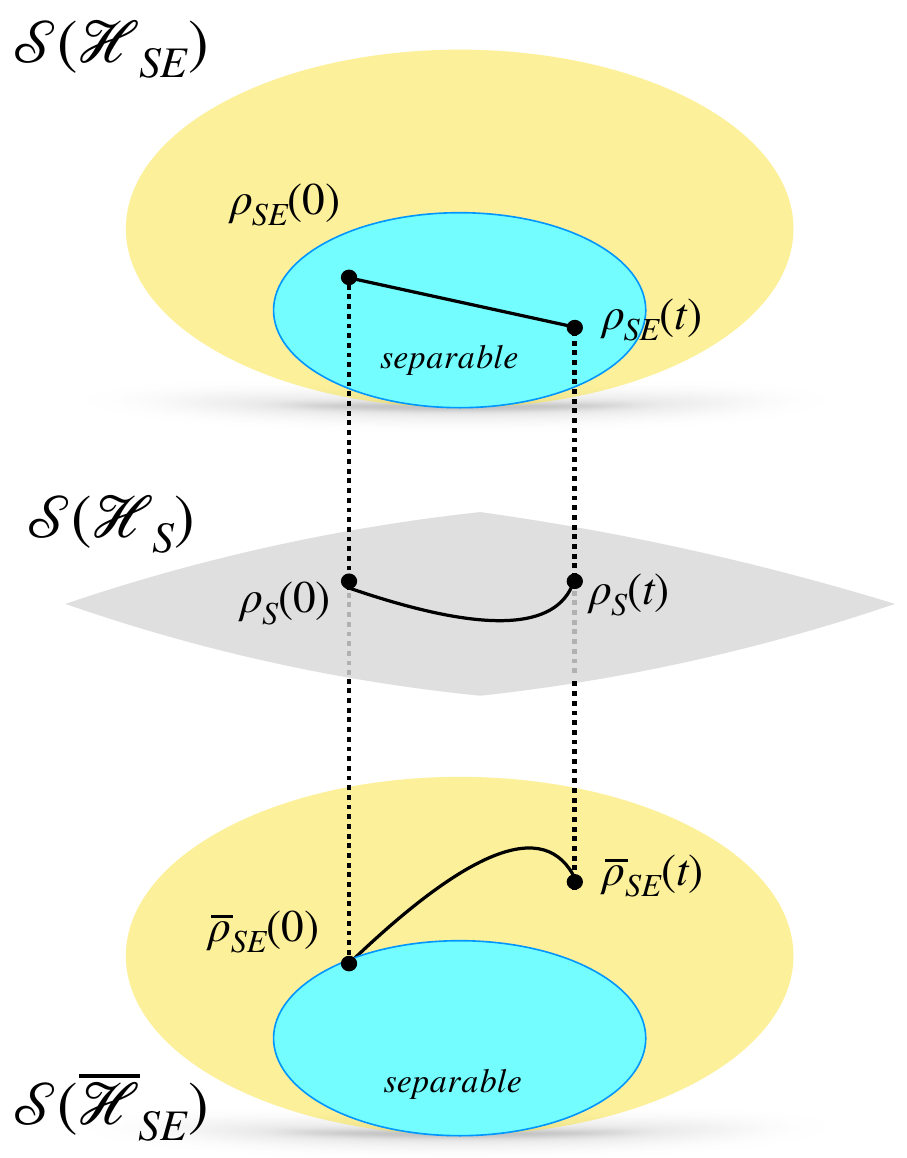}
	\caption{Sketch of two different global system-environment
	evolutions sharing the same reduced system dynamics obtained 
	by taking the partial trace as in Eq.(\ref{eq:lambda}). Note that while 
        the lower evolution involves  system-environment entangled
        states, the upper evolution takes place within the set of
        separable states. In particular, we depict it as a straight
        line to suggest that these states are actually zero-discord
        states, a subset of states that has measure zero within the
        set of separable states \cite{Ferraro2010a} (compare with
        the example in Sec.\ref{sec:secorr}).} 
	\label{fig:sk}
\end{center}
\end{figure}
More specifically, as illustrated in Fig.\ref{fig:sk},
we are going to define two different unitary evolutions associated
with the global states, respectively, $\rho_{SE}(t) \in \mathcal{S}(\mathcal{H}_{SE})$,
and $\overline{\rho}_{SE}(t) \in \mathcal{S}(\overline{\mathcal{H}}_{SE})$, which share
the same reduced system state at any time $t$,
$\rho_S(t)=\overline{\rho}_{S}(t)$. Hence, the reduced system dynamics is exactly the same 
and therefore the global exchange of information between the open system and the environment, despite the difference between the two global evolutions. 
Even more, we will show in the next section that the two global states can be characterized
by radically different kinds of correlations and environmental evolutions; 
in particular, as sketched in Fig.\ref{fig:sk}, it is possible that
one global state is a separable state at any time, while the other global state 
is entangled at (almost) any time.

\subsection{Generalized dephasing models}\label{sec:gdm}

We consider the generalized pure dephasing microscopic model \cite{Breuer2002}, 
in which a $d$-dimensional open system
interacts with an environment, according to the global Hamiltonian
$H =H_S + H_E + H_I$,
where $H_S$ and $H_E$ are the free Hamiltonians of, respectively, the open system
and the environment, 
$H_I=\sum_{n=1}^d \ket{n}\bra{n} \otimes B_n$ is the interaction Hamiltonian, with
$\left\{\ket{n}\right\}_{n=1,\ldots d}$ an orthonormal basis of
$\mathcal{H}_S$,
and $B_n = B^{\dag}_n$ are arbitrary self-adjoint operators on $\mathcal{H}_E$; 
crucially, $\left[H_S,  \ket{n}\bra{n}\right] = 0$ 
so that the free system Hamiltonian commutes with the interaction term
and the model can be solved exactly.
To do that, one can introduce the environmental interaction-picture operators
$
B_n(t) = e^{i H_E t}B_n e^{-i H_E t},
$
along with the corresponding unitaries
$
V_n(t) = T_\leftarrow \exp\left(-i \int_0^t d s B_n (s)\right),
$
where $T_\leftarrow$ is the chronological time-ordering operator.
Given the generic initial product state
$
\rho_{SE}(0)=\rho_{S}(0)\otimes \rho_{E}(0),
$
we express the initial system state with respect to the basis $\left\{\ket{n}\right\}_{n=1,\ldots d}$,
$
\rho_{S}(0)=\sum_{n,m=1}^d c_{n m}\ket{n}\bra{m},
$
while the initial environmental state with respect to its spectral decomposition,
$
\rho_{E}(0)=\sum_{\alpha} \lambda_{\alpha}\ket{\phi_\alpha}\bra{\phi_\alpha},
$
where the index $\alpha$ runs from 1 to the (possibly infinite) rank of $\rho_{E}(0)$.
Then, exploiting the linearity
of the global unitary evolution and partial trace,
the global state at time $t$ in the interaction picture with respect
to $H_S+H_E$ can be written as \cite{Breuer2002}
\begin{equation}\label{eq:global}
\rho_{SE}(t) = \sum_{n,m=1}^d \sum_{\alpha} c_{n m}  \lambda_{\alpha} 
V_n(t)\ket{n \phi_\alpha} \bra{m \phi_\alpha} V^\dag_m(t)
\end{equation}
and, by taking the partial trace over the environment (see Eq.(\ref{eq:lambda})), 
the open system state at time $t$ is
\begin{eqnarray}
\rho_{S}(t)&=&\sum_{n,m=1}^d \sum_{\alpha} c_{n m}  \lambda_{\alpha} 
\mathcal{F}_{\alpha, n, m}(t) \ket{n}\bra{m}, \label{eq:rhost}\\
\mathcal{F}_{\alpha, n, m}(t)&:=& \bra{\phi_\alpha} V^\dag_m(t) V_n(t) \ket{\phi_\alpha}.\label{eq:mathcf}
\end{eqnarray}

It is now clear how to define two global unitary evolutions
along with two initial environmental states
such that the corresponding
open system states coincide at any time $t$.
In fact, let $\left\{\lambda_\alpha, \ket{\phi_\alpha}, B_n\right\}$ and
$\left\{\overline{\lambda}_\beta, \ket{\overline{\phi}_\beta}, \overline{B}_n\right\}$
be two sets with the eigenvalues and eigenvectors of the initial environmental states
$\rho_E(0)$ and $\overline{\rho}_E(0)$
(possibly on two different Hilbert spaces $\mathcal{H}_E$ and 
$\overline{\mathcal{H}}_E$),
and the environmental interaction operators appearing in two generalized dephasing unitaries
$U_{SE}(t)$ and $\overline{U}_{SE}(t)$.
A necessary and sufficient condition to have $\rho_S(t) = \overline{\rho}_S(t)$
for any initial condition $\rho_S(0) = \overline{\rho}_S(0)$
is thus\footnote{We assume that
the free Hamiltonian $H_S$ is the same in the two cases, so that the equality among the
open system dynamics is preserved by moving back to the Schr{\"o}dinger picture.}
\begin{equation}\label{eq:cond}
\sum_\alpha \lambda_\alpha \mathcal{F}_{\alpha, n, m}(t) =
\sum_\beta \overline{\lambda}_\beta \overline{\mathcal{F}}_{\beta, n, m}(t),
\end{equation}
for any $n > m$ and $t \geq 0$, where
$\mathcal{F}_{\alpha, n, m}(t)$ and $\overline{\mathcal{F}}_{\beta, n, m}(t)$ are both
defined as in Eq.(\ref{eq:mathcf}), but with quantities referred, respectively, to the first and
to the second environment;
note that Eq.(\ref{eq:cond}) 
holds for $n>m$ if and only if it holds for $n<m$, since
$\mathcal{F}^*_{\alpha, n, m}(t)=\mathcal{F}_{\alpha, m, n}(t)$ 
and $\lambda^*_\alpha=\lambda_\alpha$.
Moreover, Eq.(\ref{eq:cond})
is satisfied automatically for $n=m$, due to the unitarity of each $V_n(t)$ and to 
the identity
$\sum_\alpha \lambda_\alpha=\sum_\beta \overline{\lambda}_\beta=1$; indeed, this
traces back to the fact that we are dealing with generalized pure dephasing dynamics,
so that the populations in the eigenbasis of the free system
Hamiltonian do not change in time. 

We stress that we did not assume any specific form of the initial states of the environments,
so that the equivalence between the open system evolutions guaranteed by Eq.(\ref{eq:cond})
does not follow from the recent equivalence theorems 
among different dynamics with initial Gaussian bosonic or fermionic environmental states \cite{Tamascelli2018,Tamascelli2019,Chen2019,Lambert2019,Nussler2020,Pleasance2020}.

\subsection{Two-level system and environment}
The condition in Eq.(\ref{eq:cond}) guarantees the equivalence
between two open system dynamics in two generalized pure-dephasing
models. To further work out analytically this equality, as well as the related quantities
referred to the global system-environment evolution, we now restrict the dimensionality of both the open system and the environment. This will also allow us to better grasp the 
physical meaning associated with Eq.(\ref{eq:cond}), relating it to the different action
of the unitaries $V_n(t)$ in Eq.(\ref{eq:global}) on populations and coherences
in the eigenbasis fixed by the initial environmental state.

First, we assume that the open system
is a two-level system, $\mathcal{H}_S = \mathbbm{C}^2$,
(with a slight change of notation we make the corresponding index $n$ run over $\left\{0,1\right\}$)
and we set $B_0=-B_1=: - B$, so that the interaction Hamiltonian
is simply the standard pure-dephasing term,
$
H_I= \sigma_z \otimes B
$,
with $\sigma_z = \ket{1}\bra{1}-\ket{0}\bra{0}$ and $B=B^\dag$ a generic self-adjoint operator on $\mathcal{H}_E$; furthermore, we also assume that
$ \left[H_E,  B\right] = 0$, so that $V(t)=e^{-i B t}$. 
As done before, we can now compare two different pure-dephasing global evolutions,
characterized by the environmental interaction operators $B$ and $\overline{B}$
and initial environmental states 
$\rho_E(0)=\sum_\alpha \lambda_\alpha \ket{\phi_\alpha}\bra{\phi_\alpha}$
and $\overline{\rho}_E(0)=\sum_\beta \overline{\lambda}_\beta \ket{\overline{\phi}_\beta}\bra{\overline{\phi}_\beta}$;
the condition in Eq.(\ref{eq:cond}) ensuring the coincidence
between the two corresponding open system dynamics
reduces to
\begin{eqnarray}
 \sum_\alpha \lambda_\alpha \bra{\phi_{\alpha}}e^{-i B t} \ket{\phi_\alpha}
 =  \sum_\beta \overline{\lambda}_\beta \bra{\overline{\phi}_{\beta}}e^{-i \overline{B} t} \ket{\overline{\phi}_\beta} 
  \label{eq:cond3}
\end{eqnarray}
for any $t\geq 0$.
The validity of Eq.(\ref{eq:cond3}) only
depends on how each pure state in the spectral decomposition of the initial 
environmental states, $\left\{\ket{\phi_\alpha}\right\}$ and $\left\{\ket{\overline{\phi}_\beta} \right\}$, 
is mapped \textit{into itself}
by the unitary operators
fixed by the environmental interaction operators, $e^{-i B t}$ and $e^{-i \overline{B} t}$.
On the other hand, the global state in Eq.(\ref{eq:global})
does depend 
on how each pure state in the spectral decomposition of the initial environmental 
states is mapped  \textit{into the other} pure states in the decomposition; 
such a dependence is precisely what is washed out
by the partial trace. This is the key mechanism guaranteeing that we can have two 
different global evolutions, with the same open system pure dephasing dynamics.

Furthermore, Eq.(\ref{eq:cond3}) can also be expressed as
\begin{equation}
 \sum_\alpha \lambda_\alpha \bra{\phi_{\alpha}} B^k \ket{\phi_{\alpha}}
 = \sum_\beta \overline{\lambda}_\beta \bra{\overline{\phi}_{\beta}} \overline{B}^k \ket{\overline{\phi}_{\beta}}
 \quad \forall k=1, \ldots;\label{exx}
\end{equation}
i.e., the equivalence of the two reduced system dynamics 
is fixed by the moments of any power of the 
interaction operators $B$ and $\overline{B}$ on the initial
environmental states $\rho_E(0)$ and $\overline{\rho}_E(0)$.
This condition might be more convenient to check \cite{Diaz2020}, especially 
if the environment has finite dimension $d_E$, so that it is enough 
to verify its validity for $k=1,\ldots d_E^2-1$,
due to the Cayley-Hamilton theorem.

Moving further, we now also restrict to the case where both pure-dephasing dynamics
are referred to two-level-system environments,  
$\mathcal{H}_E=\mathcal{H}_{\overline{E}}=\mathbbm{C}^2$,
so that $B$ and $\overline{B}$ can be seen as two
spin-$1/2$ operators associated with two different directions $\bm{\eta}$
and $\overline{\bm{\eta}}$, i.e., 
\begin{equation}\label{eq:tle}
B = g \bm{\eta} \cdot \bm{\sigma} \qquad \overline{B} = g \overline{\bm{\eta}} \cdot \bm{\sigma}, 
\end{equation}
where $\bm{\sigma}$ is the vector of Pauli matrices, $\bm{\eta}$ and $\overline{\bm{\eta}}$ are two real unit vectors,
$|\bm{\eta}|=|\overline{\bm{\eta}}|=1$, and
$g=g^*$ is the coupling constant (that is the same for the two interaction Hamiltonians).
Moreover, the two initial environmental states $\rho_E(0)$ and $\overline{\rho}_E(0)$
are fixed by two vectors $\bm{\alpha}$ and $\overline{\bm{\alpha}}$, with 
$|\bm{\alpha}|, |\overline{\bm{\alpha}}| \leqslant 1$,
according to the Bloch-ball representation \cite{Nielsen2000}
\begin{equation}\label{eq:rhoez}
\rho_{E}(0) = \frac{1}{2}\left(\mathbbm{1}+\bm{\alpha}\cdot \bm{\sigma}\right),
\quad \overline{\rho}_{E}(0) = \frac{1}{2}\left(\mathbbm{1}+\overline{\bm{\alpha}}\cdot \bm{\sigma}\right).
\end{equation}
Thus, the two open system states are the same, $\rho_S(t)=\overline{\rho}_S(t)$, 
at any time $t$ and for any 
initial condition $\rho_S(0)=\overline{\rho}_S(0)$ if and only if
\begin{equation}\label{eq:finc}
(\bm{\alpha}, \bm{\eta}) = (\overline{\bm{\alpha}}, \overline{\bm{\eta}}),
\end{equation}
as follows from Eq.(\ref{exx}) and the equality (see also Eqs.(\ref{eq:tle}) and (\ref{eq:rhoez}))
$$
\mbox{tr}\left\{\frac{1}{2}\left(\mathbbm{1}+\bm{\alpha}\cdot \bm{\sigma}\right) 
\left(\bm{\eta}\cdot \bm{\sigma}\right) \right\}
= (\bm{\alpha}, \bm{\eta}), 
$$
as well as from $B^{2k}=\overline{B}^{2k} = g^{2k}\mathbbm{1}$,
$B^{2k+1}= g^{2k+1}B$ and $\overline{B}^{2k+1}=g^{2k+1}\overline{B}$ for any $k \geq 1$.
In particular, we will focus on the case where 
$\bm{\alpha}$ and $\bm{\eta}$ are two vectors with the same
direction, but different length; explicitly,
$\bm{\alpha} = (0, 0, c)$ and $\bm{\eta} = (0, 0, 1)$,
with $c = (\bm{\alpha}, \bm{\eta})<1$. 
Note that we are excluding the value $c=1$, which would imply $\bm{\alpha} = \bm{\eta}$, 
as Eq.(\ref{eq:finc}) would then be satisfied only 
if also $\overline{\bm{\alpha}} = \bm{\overline\eta}$;
in other terms, the second pair
of (equal) vectors would simply be the rotation of the first pair
of (equal) vectors, which would correspond to a \textit{trivial} rotation on the Hilbert space of the environment.

\begin{figure}[]
  \begin{center}
    \includegraphics[width=0.45\textwidth]{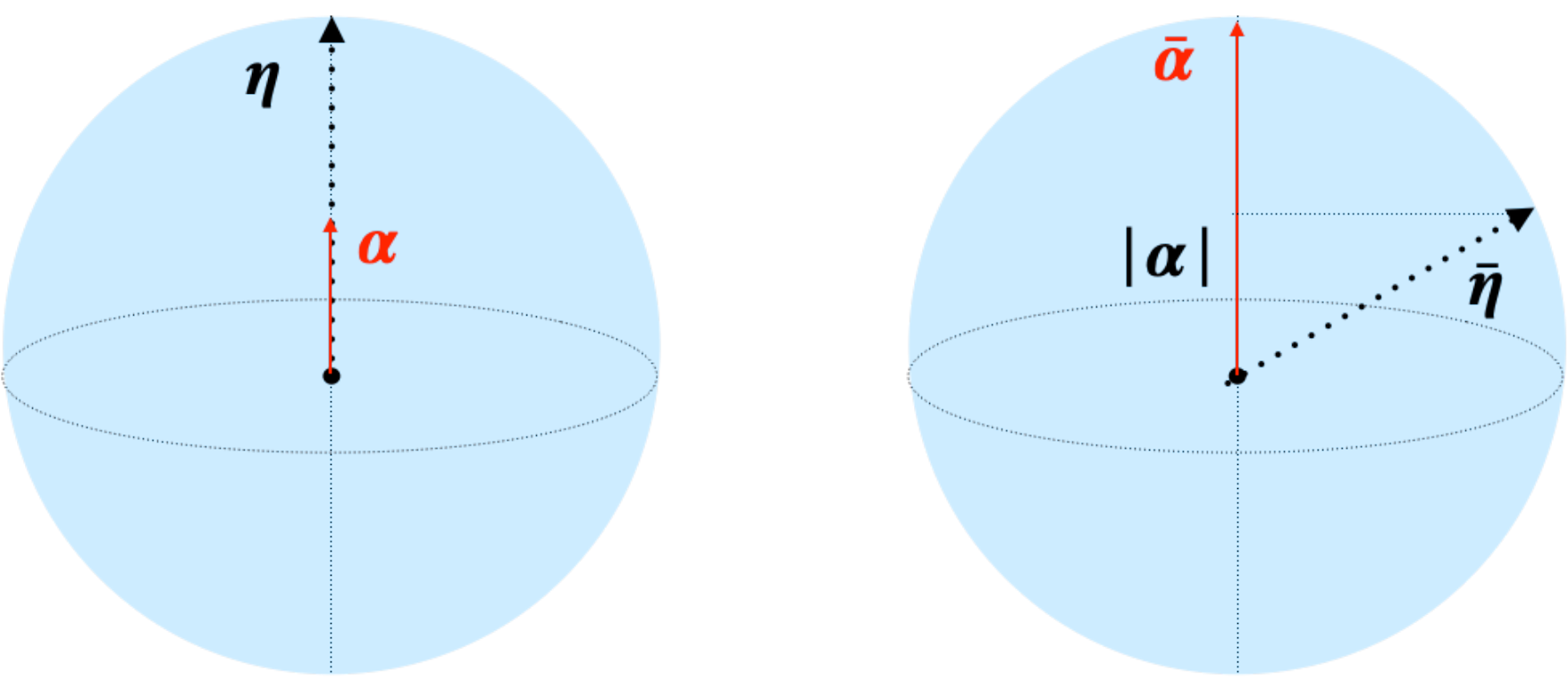}
    \caption{Geometrical meaning of the condition in
      Eq.(\ref{eq:finc}), ensuring two equal open system pure
      dephasing dynamics, in the presence of two two-dimensional
      environments.  The environmental interaction operators are fixed
      by the vectors $\bm{\eta}$ and $\overline{\bm{\eta}}$ -- see
      Eq.(\ref{eq:tle}) -- with
      $|\bm{\eta}|=|\overline{\bm{\eta}}|=1$, while the initial
      environmental states are fixed by the vectors $\bm{\alpha}$ and
      $\overline{\bm{\alpha}}$ -- see Eq.(\ref{eq:rhoez}) --
      $|\bm{\alpha}|, |\overline{\bm{\alpha}}|\leq1$; in particular,
      we consider here
      $\bm{\alpha} = (0, 0, c), \bm{\eta} = (0, 0, 1)$ and
      $|\overline{\bm{\alpha}}|=1$.}
      \label{fig:2}
  \end{center}
\end{figure}

The geometrical meaning of Eq.(\ref{eq:finc}) for the chosen $\bm{\alpha}$
and $\bm{\eta}$ is illustrated in Fig.\ref{fig:2}, under the further constraint that 
$|\overline{\bm{\alpha}}|=1$, i.e., $\overline{\rho}_E(0)$ is a pure state: in this case, 
the projection of the vector $\overline{\bm{\eta}}$ into the direction fixed
by $\overline{\bm{\alpha}}$ has to be equal to the length of the vector $\bm{\alpha}$.
In the next section, we are going to investigate the qualitatively and quantitatively
different features of the global system-environment evolutions fixed by the conditions above,
leading to the same open system dynamics.

\section{System-environment correlations, environmental states and information flow}\label{sec:secorr}
As recalled in Sect.\ref{sec:info-exchange}, the exchange of information between an
open system and its environment possibly inducing a non-Markovian evolution
is determined by the system-environment correlations and the changes in the environmental state due to the interaction between the two subsystems. The bound in Eq.(\ref{eq:bound}) 
makes this statement quantitative, via the sum of three 
contributions representing different kinds of information lying outside the open system, i.e., the 
distinguishability between the global state and the product of its marginals
for two different initial conditions and the environmental-state distinguishability related with the latter.
Starting from the two equivalent pure-dephasing dynamics introduced in the previous
section, we investigate now how qualitatively and quantitatively 
different contributions to the information related with the global evolution can result in the
same system-environment exchange of information.

\subsection{Zero-discord vs entangled global states}
First, we compare the system-environment correlations and the environmental
states in the two pure-dephasing dynamics; in the next subsection, we will finally
discuss the connection of these quantities with the system-environment information flow.

Recall that we are looking at two global evolutions where the environments are two
two-level systems, both interacting with the open system of interest via a pure dephasing
term, but fixed by two different directions, ${\bm{\eta}}$
and $\overline{\bm{\eta}}$, see Eq.(\ref{eq:tle}), and initially in two different states,
fixed by $\bm{\alpha}$ and $\overline{\bm{\alpha}}$, see Eq.(\ref{eq:rhoez}).
For the sake of concreteness, we are setting 
$\bm{\alpha} = (0, 0, c), \bm{\eta} = (0, 0, 1)$ and $\overline{\bm{\alpha}} = (0, 0, 1)$, with $c<1$.
This means that the
initial environmental state is the mixed state
$\rho_E(0)= \frac{1+c}{2}\ket{1}\bra{1}+\frac{1-c}{2}\ket{0}\bra{0}$
in the first model and the pure state
$\overline{\rho}_E(0)= \ket{1}\bra{1}$ in the second model.
Finally, the validity of Eq.(\ref{eq:finc})
guaranteeing the equivalence between the two open system dynamics, 
$\rho_S(t)=\overline{\rho}_S(t)$ for every $\rho_S(0)=\overline{\rho}_S(0)$ and $t \geq 0$,
implies that $\overline{\bm{\eta}}=(\sqrt{1-c^2-d^2},d,c)$, for any $-1\leqslant d \leqslant 1$. 
Note that due to the invariance of the trace distance under unitary operations we can set $d=0$ without loss of generality.

\begin{figure}[ht!]
\begin{center}
   \includegraphics[width=0.45\textwidth]{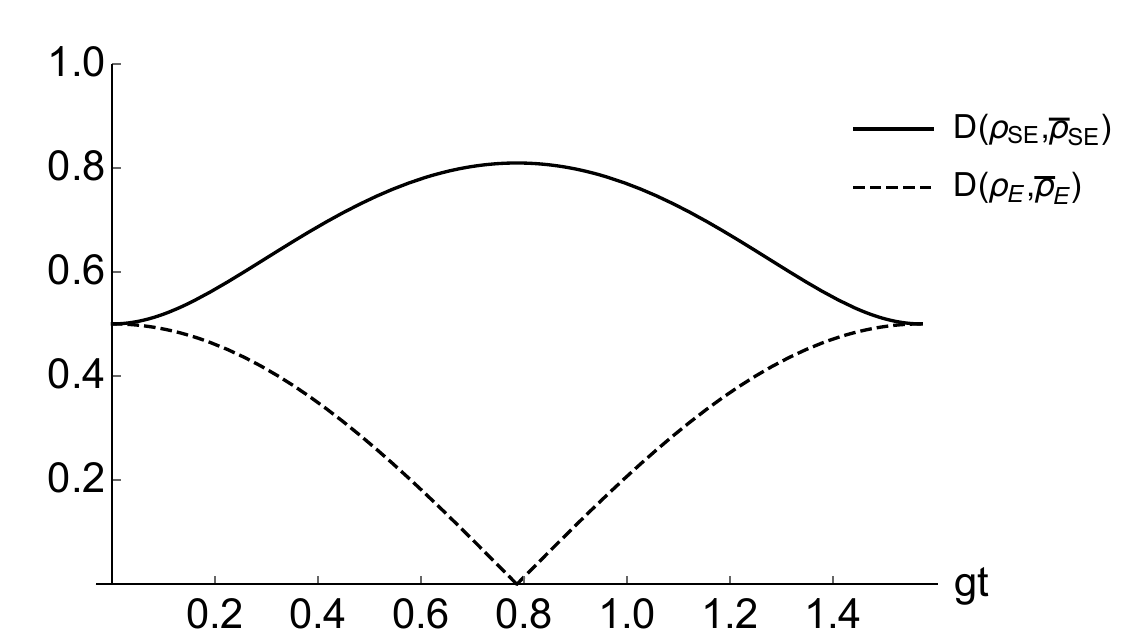}
   \caption{Time evolution of the trace distance between the 
     global states, $D(\rho_{SE}(t),\overline{\rho}_{SE}(t))$ (solid
     line), and the environmental states,
     $D(\rho_{E}(t),\overline{\rho}_{E}(t))$ (dashed line), for the
     two pure dephasing models fixed by the vectors $({\bm{\alpha}},
        {\bm{\eta}})$ and $(\overline{\bm{\alpha}},
        \overline{\bm{\eta}})$ respectively --
	see Fig.\ref{fig:2}. In both cases the system starts in the pure
     state $\ket{\psi_+}={1}/{\sqrt{2}}\left(\ket{0}+\ket{1}\right)$, while the initial states of the 
     environments are given in Eq.(\ref{eq:rhoez}).
     We take $\bm{\alpha} = (0, 0, 0)$ and $\bm{\eta} = (0, 0, 1)$,
together with $\overline{\bm{\alpha}} = (0, 0, 1)$
	and 	$\overline{\bm{\eta}}=(1, 0, 0)$.}
	\label{fig:4}
\end{center}
\end{figure}

Evaluating the two global states via Eq.(\ref{eq:global}), the difference in
their system-environment correlations appears immediately clear, showing
that we are indeed in the situation illustrated in Fig.\ref{fig:sk}\footnote{Eq.(\ref{eq:global}) refers to the global
state in the interaction picture with respect to $H_S+H_E$; on the other hand, the latter is related to
the state in the Schr{\"o}dinger picture via the factorized 
unitary operator $e^{-i H_S t} \otimes e^{-i H_E t}$,
which does not affect the system-environment correlations.
For future convenience we also note that the three terms at the r.h.s. of Eq.(\ref{eq:bound})
do not change when moving from the interaction to the Schr{\"o}dinger picture or viceversa,
due to the invariance of the trace distance under unitary operations.
Finally, the comparison between states related to different dynamics performed
in Fig.\ref{fig:4} is the same in the interaction and Schr{\"o}dinger picture, due to the specific
choice of the initial states.
}.
In the first model,
the global state at time $t$ is
\begin{eqnarray}
\!\! \!\! \!\! \!\!\rho_{SE}(t) &=& \frac{1+c}{2}  \left({\begin{array}{cc}
   c_{11} & c_{1 0}e^{-2 i g t}  \\
   c_{0 1}e^{2 i g t} & c_{00}
  \end{array} } \right) \otimes\ket{1}\bra{1}+ \label{eq:glv1} \nonumber\\
  && \frac{1-c}{2}  \left({\begin{array}{cc}
   c_{11} & c_{1 0}e^{2 i g t}  \\
   c_{0 1}e^{-2 i g t} & c_{00}
  \end{array} } \right) \otimes\ket{0}\bra{0}, 
\end{eqnarray}
which is a zero-discord state, indicating
the classical nature of the correlations between the open system and the environment \cite{Ollivier2001,Henderson2001,Modi2012}; zero-discord states
are a proper subset of the set of separable states.
On the other hand, in the second model the global state at time $t$ can be written as
\begin{eqnarray}
\overline{\rho}_{SE}(t) &=& c_{11} \ket{1}\bra{1}\otimes  
\left({\begin{array}{cc}
   |\ell_t|^2 & \ell_t^* \kappa_t  \\
   \ell_t \kappa_t^* & |\kappa_t|^2  
       \end{array} } \right)+\nonumber \\
  &&c_{00}\ket{0}\bra{0}\otimes 
\left({\begin{array}{cc}
   |\ell_t|^2 & -\ell_t \kappa_t  \\
   -\ell_t^* \kappa_t^* &|\kappa_t|^2  
  \end{array} } \right)+\\
&&c_{10} \ket{1}\bra{0} \otimes \left({\begin{array}{cc}
   \ell_t^{*2} & -\ell_t^* \kappa_t  \\
   \ell_t^* \kappa_t^* & -|\kappa_t|^2  
  \end{array} } \right)+h.c.
, \nonumber
\end{eqnarray}
where $h.c.$ denotes the Hermitian conjugate of the term at its own left and
\begin{equation}
 \ell_t = \cos(g t)+i c \sin(g t); \quad
\kappa_t =i \sqrt{1-c^2} \sin(g t).\label{eq:glv2}
\end{equation}  
This state is easily shown to be an entangled state at almost every time $t$,
e.g., by means of the partial transposition criterion
\cite{PhysRevLett.77.1413, Horodecki1996}.
More generally, any pure-dephasing evolution will generate entanglement
between the two-level system and its (generic) environment
if and only if the initial state of the environment does not commute with the environmental unitary interaction
operator  $V(t)$ (see the definition at the beginning of Sec.\ref{sec:gdm}) 
\cite{Roszak2015}.
In addition, we stress that two unitary dilations for the same pure dephasing CPTP map, one associated
with a global entangled state and one with a zero-discord state have been derived
in \cite{Costa2016}.

Actually, one can also quantify explicitly the amount of entanglement of $\overline{\rho}_{SE}(t)$
by using the concurrence, according to \cite{Wootters1998}
\begin{equation}\label{eq:conc}
\mathcal{C}[\overline{\rho}_{SE}(t)] = \max\left\{0,\lambda_1(t)-\lambda_2(t)
-\lambda_3(t)-\lambda_4(t)\right\},
\end{equation}
where $\lambda_1(t)\geq\lambda_2(t)\geq\lambda_3(t)\geq\lambda_4(t)$
are the square root of the eigenvalues of 
$\overline{\rho}_{SE}(t)(\sigma_y\otimes \sigma_y)\overline{\rho}^*_{SE}(t)(\sigma_y\otimes \sigma_y)$, with $\sigma_y = -i \ket{0}\bra{1}+i \ket{1}\bra{0}$
and $\rho^*$ the complex conjugate of $\rho$. 
In fact, the interaction between the open
system and the environment leads to the presence of entanglement for any time $t>0$,
apart from isolated instants of time, as quantified by $\mathcal{C}[\overline{\rho}_{SE}(t)]$; noticeably, maximally entangled states, for
which the value of concurrence is equal to 1, can be generated by the global evolution.

The difference between the two global evolutions is further illustrated in Fig.\ref{fig:4},
where we report the evolution of the trace distance between the two corresponding global
states, $D(\rho_{SE}(t),\overline{\rho}_{SE}(t))$, and the two environmental states 
$D(\rho_{E}(t),\overline{\rho}_{E}(t))$.
At the initial time the two quantities coincide, since both initial system-environment states
are product states. Then,
while $D(\rho_{SE}(t),\overline{\rho}_{SE}(t))$ takes values
greater or equal to its initial value,
$D(\rho_{E}(t),\overline{\rho}_{E}(t))$ oscillates between
its initial value and zero; the contractivity of the trace distance under CPTP maps
implies that $D(\rho_{E}(t),\overline{\rho}_{E}(t)) \leqslant D(\rho_{SE}(t),\overline{\rho}_{SE}(t))$.
Interestingly, we also note that when the global-state
distinguishability increases the environmental-state distinguishability decreases and viceversa,
so that when the two environmental states 
coincide, $D(\rho_{E}(t),\overline{\rho}_{E}(t))=0$, the two global states have reached
their maximum value of distinguishability, which is then fully due
to the different correlations in the two global states.

\subsection{Different contributions to the system-environment exchange of information}
We have thus seen that different evolutions of the global states 
and the system-environment correlations can still lead
to the same open system dynamics, meaning in particular that the 
quantum or classical nature of the system-environment correlations is not crucial for the 
presence of memory effects in the dynamics at hand \cite{Pernice2011,Pernice2012,Mazzola2012,Smirne2013,Campbell2019}.
\begin{figure}[hb!]
\begin{center}
	    \includegraphics[width=0.45\textwidth]{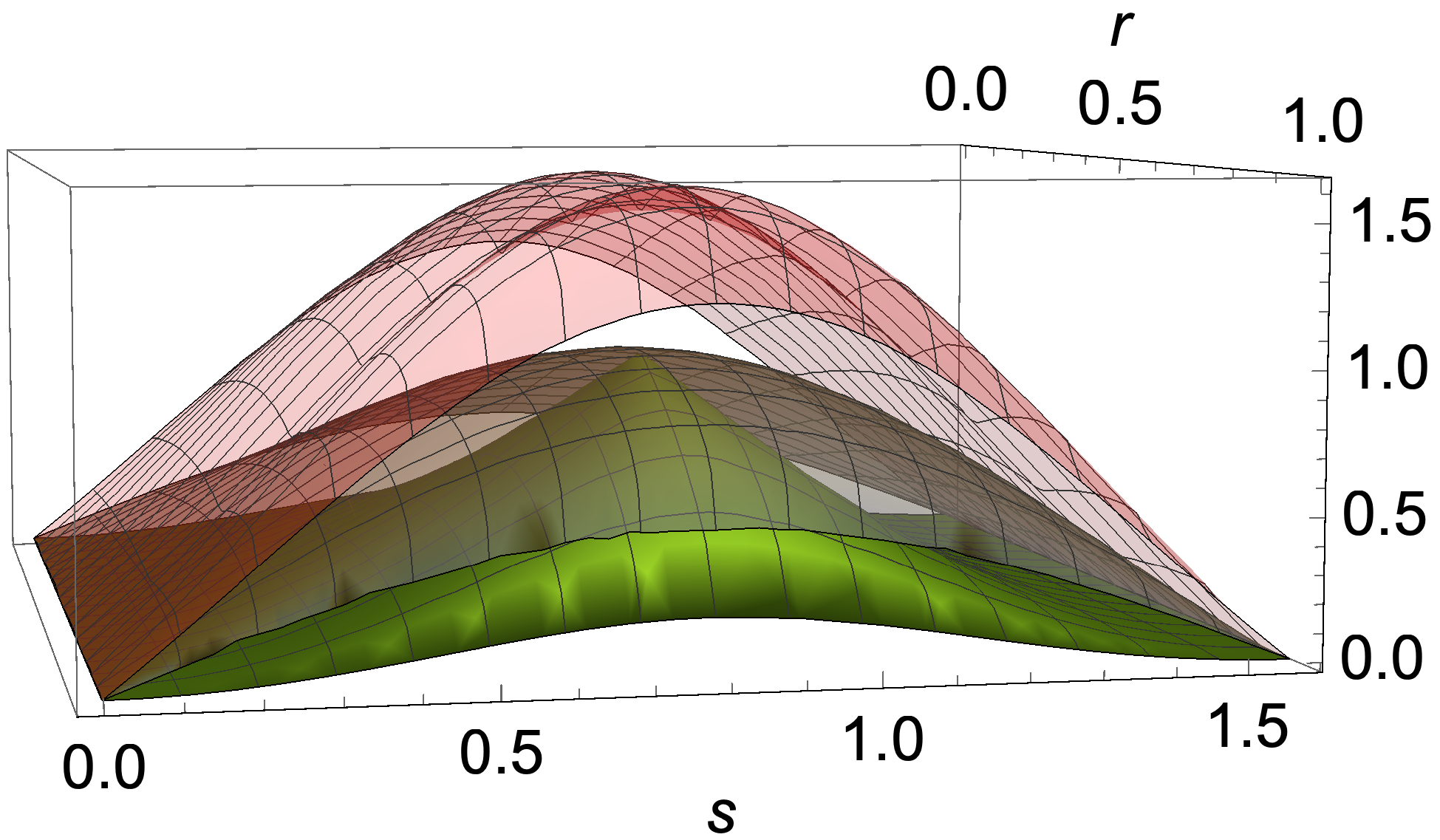}
	\caption{Trace distance variation $\Delta_S(t,s)$ (green
          surface) for $t=\pi/2$ as a function of the time $s$
          and the parameter $r$ determining the initial condition.
          The pair of initial states for the system is given by the
          pure states $\rho^1_S(0)= \ket{\psi_+}\bra{\psi_+}$ and
          $\rho^2_S(0)=\ket{\psi_-^{r}}\bra{\psi_-^{r}}$, with
          $\ket{\psi_-^{r}}=\left(r\ket{0}-\sqrt{1-r^2}\ket{1}\right)$.
          The black and red meshed transparent surfaces correspond to
          the bounds $I_{SE}(s)$
          and $\overline{I}_{SE}(s)$ respectively, according to Eqs.(\ref{eq:bound}) and
          (\ref{eq:bound2}). The other parameters are as in
          Fig.\ref{fig:4}.}
    \label{fig:5}
  \end{center}
\end{figure}

\begin{figure}[hb!]
\begin{center}
	    \includegraphics[width=0.45\textwidth]{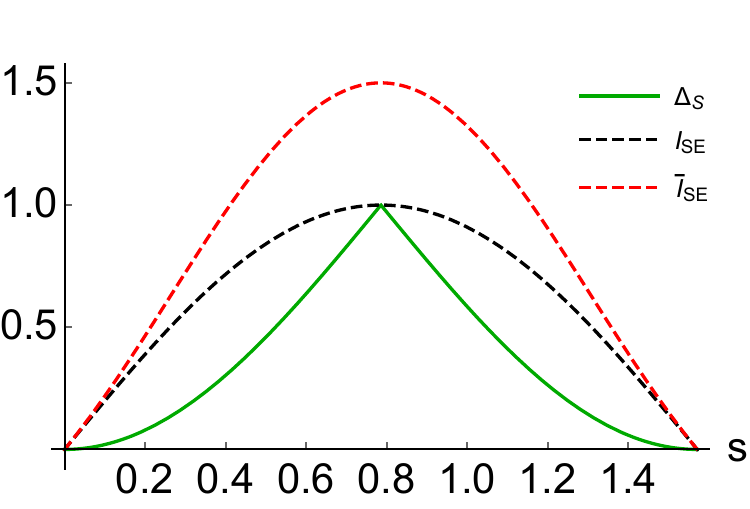}
	\caption{Section of Fig.\ref{fig:5}, for the value $r=1/\sqrt{2}$. The
      plot shows the bounds $I_{SE}(s)$ and $\overline{I}_{SE}(s)$
      (dashed black and red lines respectively) together with the trace
      distance variation $\Delta_S(t,s)$  for $t=\pi/2$ (solid green line). It
      clearly appears saturation of the upper bound for the considered
      pair of initial system states.}
    \label{fig:6}
  \end{center}
\end{figure}

We now move one step forward and use the three contributions at the r.h.s.
of the bound in Eq.(\ref{eq:bound}) to quantify the different kinds of information lying
outside the open system and their relation with the system-environment information flow. 
In Fig.\ref{fig:5},
we depict with a green surface the open system trace distance variation
$\Delta_S(\pi/2,s)$ defined in Eq.(\ref{eq:variation}), which by construction
is the same for the two pure-dephasing models we are dealing with;
indeed $\Delta_S(\pi/2,s)$ is always larger than zero for the chosen time
interval, in accordance with
the strong non-Markovian character of the open system pure dephasing dynamics due to the interaction
with a two-level system environment.
The overall amount of information contained in the system-environment
correlations and environmental-state distinguishability, as quantified via the sum of
the three contributions at the r.h.s. of
Eq.(\ref{eq:bound2}), is represented by the meshed trasparent black and
red surfaces. In Fig.\ref{fig:6}, we consider a section of 
Fig.\ref{fig:5}
corresponding to a fixed pair of initial system states.
It clearly appears that the sum of system-environment correlations and environmental-state
distinguishability in the model characterized by the presence of entanglement exceeds
the corresponding sum for the classically-correlated model, so that
the bound on the open system trace distance given by
Eq.(\ref{eq:bound}) is tighter in the latter case and one can
consider a choice of initial pure states such that the bound is
actually saturated at some intermediate point of time.
This is exactly the choice we have made in Fig.\ref{fig:6}.

Despite the different amount of
information associated with system-environment correlations and environmental states
distinguishabilities in the two models, 
the open system dynamics that results after averaging out the environmental degrees of freedom is exactly the same. The mentioned differences do not affect
in any way the information exchange between the open system and the environment
and therefore the relevance of memory effects in the open system dynamics, as quantified
by the magnitude of the trace distance revivals.
\begin{figure}[]
\begin{center}
	\includegraphics[width=.45\textwidth]{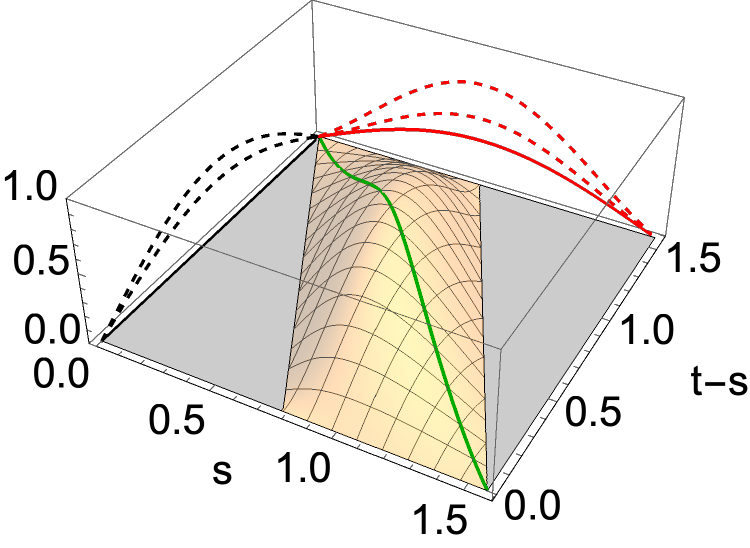}
	\caption{Trace distance variation
$\Delta_S(t,s)$, see Eq.(\ref{eq:variation}), as a function of $s$ and $t-s$
for the two pure-dephasing models fixed by $(\bm{\alpha}, \bm{\eta})$ and 
$(\overline{\bm{\alpha}}, \overline{\bm{\eta}})$ -- see Fig.\ref{fig:2}.
The green solid line in the transparent plane corresponds to fixing $t=\pi/2$.
The black and red lines in the background  correspond to the three
distinct contributions of  $I_{SE}(s)$ (left,black) and
$\overline{I}_{SE}(s)$ (right,red) respectively.
For the model fixed by $(\bm{\alpha}, \bm{\eta})$ (left) the dashed
lines correspond to the total amount of system-environment correlations as a function of the time $s$, while the solid line
depicts the distinguishability between the two environmental
marginals, which always remain the same.
The same quantities are plotted for the model fixed by
$(\overline{\bm{\alpha}}, \overline{\bm{\eta}})$ (right). It clearly
appears that in this case the information initially contained in the open system state is 
later stored also in the environmental degrees of freedom.
For both dynamics the initial reduced system states are
$\rho^1_S(0)= \ket{\psi_+}\bra{\psi_+}$,
$\rho^2_S(0)=\ket{\psi^{0.4}_{-}}  \bra{\psi^{0.4}_{-}} $, while the other parameters 
are as in Fig.\ref{fig:4}.
}
	\label{fig:7}
\end{center}
\end{figure}
Interestingly, relevant differences can be observed also if we look at each
of the three contributions at the r.h.s.
of Eq.(\ref{eq:bound}) individually.  The latter are represented by
the lines on the two planes in the background of Fig.\ref{fig:7}, where each plane
refers to one of
the two pure-depashing models, while the 3D plot depicts the identical trace distance
variation $\Delta_S(t,s)$ for the two models, as a function of $s$ and $t-s$. We can see that for the first dynamics (black lines) the environmental states remain the same for both chosen initial states. On the other hand, in the second model (red lines) the environmental states do depend on the initial open system states, showing that in this case the environmental degrees of freedom have an important role in storing information that was previously in the reduced quantum system. In addition, the amount of information 
contained in the correlations
for both initial conditions differs significantly in the two microscopic models.
Hence, Fig.\ref{fig:7} yields a direct illustration of how different contributions to the information content
outside the open system -- being in the system environment correlations or in environmental-state
distinguishability -- can result in the very same flow of information towards the reduced system.

\section{Conclusions and outlook}\label{sec:concl}
\label{sec:conclusions-outlook}
In this paper, we have investigated the microscopic origin of the exchange of information
between an open quantum system and its environment. To do so, we have considered two
generalized pure dephasing microscopic models, with different environmental states and system-environment interaction terms, leading to the same reduced system dynamics. 
In this way, we have shown how quantitatively and even qualitatively different features of the 
information contained in system-environment
correlations and environmental states might well result in the same flow of information towards the open system, implying the same increase in the trace distance and thus the same amount of non-Markovianity
in the dynamics. In particular, the first model is characterized by classical system-environment
correlations (that is, the global state has always zero discord), while the second 
generates entangled global states at almost any time; 
in addition, for a specific choice of the initial conditions, in the first
model the environmental states do not depend on the open system initial state, while 
in the second model
significant information is contained in the environmental-state distinguishability.

Indeed, it will be important to investigate to which extent the results we obtained
in the presence of pure dephasing can be extended to more complex system-environment
microscopic models. As a first step, one could consider higher dimensional open systems
where the equality in Eq.(\ref{eq:cond}) still ensures the equivalence between the reduced dynamics
of different models;
as an example, dealing with multi-qubit open systems \cite{Costa2016} might also allow us to study the interplay between the system-environment exchange of information and the correlations within the open system itself. More in general, to go beyond (generalized) pure dephasing models, one could 
resort to approximate solutions
of the dynamics and to numerical methods. While giving access to a much larger
class of systems, this kind of approaches would however inevitably weaken 
the exact equivalence between the reduced dynamics of different microscopic models, 
which is instead one of the main motivations of our analysis. 
An alternative path could then be to compare the open system dynamics due to an overall
unitary evolution with the dynamics induced by a non-unitary evolution,
by means of the so-called Lindbladian embedding methods
\cite{Imamoglu1994,Garraway1997,Tamascelli2018,Tamascelli2019,Chen2019,Lambert2019,Somoza2019,Nussler2020,Pleasance2020,Luchnikov2020}. Here, the global system is complex enough to account
for a large variety of realistic models and would possibly require the use of approximated or numerical 
techniques to evaluate global system-environment quantities; on the other hand, crucially, the equivalence between the reduced
dynamics of different models would still be guaranteed a-priori in an exact way.

In conclusion, we hope that our results, along with future investigation
initiated by them, will provide a useful reference point to 
understand the general physical mechanisms ruling the origin of non-Markovianity, 
identifying those global features that unavoidably impact on the
behavior of proper quantifiers of the information accessible via the
open quantum system evolution.

\section*{Acknowledgements}
NM would like to thank Walter T. Strunz for getting her interested in the topic.
NM acknowledges funding by the Alexander von Humboldt Foundation in the form of a Feodor-Lynen
Fellowship. All authors acknowledge support
from the UniMi Transition Grant H2020.

\end{document}